\documentclass[conference]{IEEEtran}
\usepackage{array}
\usepackage{algorithm, algpseudocode}
\IEEEoverridecommandlockouts
\usepackage{tikz}
\usetikzlibrary{positioning, shapes.geometric, arrows.meta}
\usepackage{cite}
\usepackage{multirow} 
\usepackage{orcidlink}
\usepackage{amsmath,amssymb,amsfonts}
\newcounter{examplecounter}
\newenvironment{example}{\begin{quote}%
    \refstepcounter{examplecounter}%
  \textbf{Example \arabic{examplecounter}}%
  \quad
}{%
\end{quote}%
}
\usepackage{graphicx}
\usepackage{rotating}
\usepackage{textcomp}
\usepackage{xcolor}
\def\BibTeX{{\rm B\kern-.05em{\sc i\kern-.025em b}\kern-.08em
    T\kern-.1667em\lower.7ex\hbox{E}\kern-.125emX}}
\begin{document}

\title{Enhancing the conformal predictability of context-aware recommendation systems by using Deep Autoencoders}

\author{
    \IEEEauthorblockN{
        Saloua Zammali\IEEEauthorrefmark{1}\orcidlink{0009-0008-5955-9534}, 
        Siddhant Dutta\IEEEauthorrefmark{2}\orcidlink{0009-0000-5120-7114}, 
        Sadok Ben Yahia\IEEEauthorrefmark{3}\IEEEauthorrefmark{4}\orcidlink{0000-0001-8939-8948}        
    }
    \IEEEauthorblockA{        
        \IEEEauthorrefmark{1}Faculty of Science of Tunis, University of Tunis El-Manar, Tunis, Tunisia\\
        \IEEEauthorrefmark{2}SVKM's Dwarkadas J. Sanghvi College of Engineering, Mumbai, India\\
        \IEEEauthorrefmark{3}The Maersk Mc-Kinney Moller Institute, University of Southern Denmark, Sønderborg, Denmark\\
        \IEEEauthorrefmark{4}Department of Software Science, Tallinn University of Technology, Tallinn, Estonia\\        
        Email: saloua.zammali@fst.utm.tn, siddhant.dutta180@svkmmumbai.onmicrosoft.com, say@mmmi.sdu.dk
    }
}

\maketitle

\begin{abstract}
In the field of Recommender Systems (RS), neural collaborative filtering represents a significant milestone by combining matrix factorization and deep neural networks to achieve promising results. Traditional methods like matrix factorization often rely on linear models, limiting their capability to capture complex interactions between users, items, and contexts. This limitation becomes particularly evident with high-dimensional datasets due to their inability to capture relationships among users, items, and contextual factors. Unsupervised learning and dimension reduction tasks utilize autoencoders, neural network-based models renowned for their capacity to encode and decode data. Autoencoders learn latent representations of inputs, reducing dataset size while capturing complex patterns and features.  In this paper, we introduce a framework that combines neural contextual matrix factorization with autoencoders to predict user ratings for items. We provide a comprehensive overview of the framework’s design and implementation. To evaluate its performance, we conduct experiments on various real-world datasets and compare the results against state-of-the-art approaches. We also extend the concept of conformal prediction to prediction rating and introduce a Conformal Prediction Rating (CPR). For RS, we define the nonconformity score, a key concept of conformal prediction, and demonstrate that it satisfies the exchangeability property.

\end{abstract}


\section{Introduction}
Recommender systems predict user ratings for specific items using a formula that maps the relationship between users, items, and ratings:
$R: \text{{user}} \times \text{{item}} \rightarrow \text{{rating}}$.
To further enhance recommendation performance, Context-Aware Recommender Systems (CARS) have gained considerable attention for their ability to incorporate contextual information. Unlike traditional approaches that rely solely on user-item interactions, CARS aim to refine the rating function by considering user, item, and context:
$R: \text{{user}} \times \text{{item}} \times \text{{context}} \rightarrow \text{{rating}}$
By integrating contextual data, CARS provide more accurate and relevant recommendations, addressing the limitations of traditional recommender systems \cite{ZammaLi2021}.
Matrix Factorization (MF) techniques are among the most widely used and popular models in collaborative filtering\cite{DridiZAA20}. The core concept of matrix factorization is to extract latent features, or embeddings, 
from the observed data in matrix $R$. Since users don't rate all items, the matrix is often very sparse, with only a small portion of the matrix cells filled with ratings while the majority remain empty.

In this paper, we present an innovative deep learning recommendation model that incorporates contextual information. Building upon the neural collaborative filtering and neural matrix factorization models proposed by He et al. \cite{He2017NeuralCF}, which leverage deep neural networks to learn nonlinear functions of user-item interactions rather than the fixed linear inner product used in traditional matrix factorization, our approach extends these models to encompass complex interactions among users, items, and contexts. Our model demonstrates how integrating contextual information enhances the accuracy of rating predictions by augmenting the neural recommendation models.

This paper is organized as follows: Section \ref{RelatedWorks} and \ref{Neuracollab} provides a comprehensive review of state-of-the-art research in recommender systems, highlighting various approaches, techniques, and algorithms used in previous studies. Section \ref{Model} introduces our proposed framework for a neural contextual matrix factorization recommender system utilizing autoencoders. Section \ref{Experiments} details the experiments conducted to evaluate the effectiveness and performance of different models within our framework. Finally, Section \ref{Conclusions} summarizes the key findings and contributions of our research, and discusses potential directions for future work, highlighting areas that warrant further investigation and improvement.

\section{Related Works
\label{RelatedWorks}}

\subsection{Traditional Matrix Decomposition Baselines}
Matrix factorization is a widely used collaborative filtering technique in recommender systems. The classic approach predicts ratings \( r_{ij} \) by computing the dot product of user embedding \( u_i \) and item embedding \( v_j \):
\begin{equation}
r_{ij} = u_i \cdot v_j
\end{equation}

However, rating values can vary significantly across users and items, and their interpretation can differ. For example, User A, who typically gives high ratings with an average of 4, might view a rating of 3 as indicating a low preference for an item. In contrast, User B, who usually rates items lower with an average of 2, might consider a rating of 3 as the highest possible rating. Consequently, the meaning of a rating value can vary greatly between users. Similarly, different items may receive consistently negative or positive ratings.

To account for these variations, researchers have incorporated user and item biases into matrix factorization models, enhancing performance \cite{koren2009matrix}. Biased matrix factorization extends the basic model by adding two bias terms: one for the user ($b_i$) and one for the item ($b_j$). These biases capture the tendencies of users and items toward certain rating levels, improving the rating prediction equation:
\begin{equation}
r_{ij} = u_i \cdot v_j + b_i + b_j
\end{equation}

Several extensions to matrix factorization have been developed, each addressing specific challenges. Singular Value Decomposition (SVD) \cite{Klema1980TheSV}, for instance, decomposes the rating matrix ($R$) into three submatrices. Probabilistic Matrix Factorization (PMF) \cite{Salakhutdinov2007ProbabilisticMF} introduces a probabilistic framework to model uncertainties in rating predictions. Non-negative Matrix Factorization (NMF or NNMF)\cite{Wang2013NonnegativeMF} imposes constraints that require all elements of the factor matrices to be non-negative, offering a different perspective on the factorization process.
\subsection{Autoencoder and Deep Autoencoder Models}
\label{AutoencoderRelatedWorks}

Sedhain et al. \cite{Sedhain2015AutoRecAM} introduced AutoRec, an autoencoder-based framework for collaborative filtering. AutoRec demonstrated improved performance over traditional methods like Biased Matrix Factorization (MF) on popular datasets such as MovieLens and Netflix, despite its simplicity.

AutoRec consists of two models: \textbf{U-AutoRec} and \textbf{I-AutoRec}.

\textbf{U-AutoRec} operates on user-based data. It processes a sparse vector of user ratings for all available items. Specifically, for a matrix \( R \) representing \( n \) users and \( m \) items, U-AutoRec takes as input a vector \( r^{(u)} = (R_{u1}, \ldots, R_{um}) \), where each element \( R_{uj} \) contains the user’s rating for item \( j \), or 0 if the item was not rated. This input is mapped to a lower-dimensional latent space and then reconstructed in the output space. The model’s objective is to predict the missing ratings (i.e., the zero entries in the input vector). The loss function used for training is the Mean Masked Squared Error (MMSE), which focuses on minimizing the reconstruction error for observed ratings while ignoring unobserved (missing) ratings. The loss function is formulated as follows:

\[
L = \sum_{u=1}^{n} \left\| r^{(u)} - h(r^{(u)}; \theta) \right\|_\mathcal{O}^2 + \frac{\lambda}{2} \cdot \left( \|W\|_F^2 + \|V\|_F^2 \right)
\]

In this equation, \( \| \cdot \|_{\mathcal{O}}^2 \) restricts the focus to observed ratings (non-zero entries), and \( h(r^{(u)}; \theta) \) represents the reconstruction of \( r^{(u)} \) based on the learned parameters \( \theta \):

\[
h(r^{(u)}; \theta) = f(W \cdot g(Vr^{(u)} + \mu) + b)
\]

\textbf{I-AutoRec}, on the other hand, processes item-based data. It takes a vector of ratings for a specific item, given by all users, and applies a similar process to predict missing ratings.
Building on AutoRec, Kuchaiev \cite{Kuchaiev2017TrainingDA} proposed a \textbf{deep autoencoder model} to improve generalization and performance. This deep autoencoder uses a technique called dense refeeding during training, which makes it more robust than shallower models.
The training process for the deep autoencoder involves the following steps:
\begin{enumerate}
    \item \textbf{Initial Forward Pass:} Given a sparse input \( x \) (e.g., a vector of user ratings), the model computes a dense output \( f(x) \) and calculates the loss using the same MMSE function described in Equation 1.
    \item \textbf{Backward Pass:} The model computes gradients and updates the weights using backpropagation.
    \item \textbf{Dense Refeeding:} The output from the first forward pass, \( f(x) \), is treated as a new input, and a second forward pass is performed to compute \( f(f(x)) \), which is now a fully dense vector (all values are non-zero). The loss is again computed using this fully dense vector.
    \item \textbf{Second Backward Pass:} Gradients are recalculated and the weights are updated again based on the second forward pass.
\end{enumerate}
\section{Other Neural Collaborative Filtering Models \label{Neuracollab}}

Although matrix factorization techniques have yielded impressive results, they often assume linear relationships between users and items, which may not effectively capture complex real-world interactions.

To address this, Xue et al. \cite{Xue2017DeepMF} proposed the Deep Matrix Factorization (DeepMF) model. DeepMF inputs two sparse vectors of ratings: \( U_i \) for user \( i \) and \( V_j \) for item \( j \). These vectors are processed through multiple neural network layers, and the final output is obtained using a cosine similarity layer:

\begin{equation}
\text{Cosine Similarity}({p}_i, {q}_j) = \frac{{\mathbf{p}_i^T \mathbf{q}_j}}{{\lVert \mathbf{p}_i \rVert \lVert \mathbf{q}_j \rVert}}
\label{eq 12}
\end{equation}

He et al. \cite{He2017NeuralCF} introduced the Neural Collaborative Filtering (NCF) framework, which enhances traditional collaborative filtering by employing neural networks. NCF processes sparse binary user and item vectors through an embedding layer to learn meaningful representations. Additionally, He et al. proposed the Neural Matrix Factorization (NeuMF) framework, which combines General Matrix Factorization (GMF) and Multilayer Perceptrons (MLP). GMF learns low-order features through element-wise products of embeddings, while MLP learns high-order features through dense neural networks. The NeuMF framework integrates outputs from both GMF and MLP to improve prediction accuracy.

Bobadilla et al. \cite{Bobadilla2022NeuralCF} proposed a Neural Collaborative Filtering Classification Model that diverges from DeepMF and NCF by using only user and item IDs. These IDs are encoded into vectors through separate embedding layers. Rather than predicting a single rating, the model outputs a vector of probabilities for each rating value, selecting the highest probability rating as the final prediction. This multiclass classification approach offers a more detailed prediction by considering various rating possibilities.

\subsection{Deep Context-Aware Models}

Unger et al. \cite{Unger2020ContextAwareRB} introduced an extension to the Neural Matrix Factorization (NeuMF) framework that incorporates contextual information. This extension builds on He et al.'s original NeuMF framework by integrating contextual factors such as time, location, or user context into the recommendation process. The framework offers three methods for representing contextual factors. The first method, explicit representation, maintains the original context, where nominal contexts are encoded using one-hot encoding, and ordinal and quantitative contexts are normalized. The second method, unstructured latent representation, uses an autoencoder to capture complex contextual relationships. The third method, structured latent representation, employs a hierarchical tree to leverage hierarchical dependencies among contextual factors. This results in three distinct models: the Explicit Context-Aware Model (ECAM), the Unstructured Context-Aware Model (UCAM), and the Hierarchical Context-Aware Model (HCAM).

Zheng et al. \cite{Zheng2022AFO} further advanced the work of Unger et al. by proposing a family of Neural Contextual Matrix Factorization (NeuCMF) models. This framework replaces the Generalized Matrix Factorization (GMF) tower with three distinct towers: UI, UC, and IC. The UI tower, analogous to the GMF tower, captures user-item interactions. The UC tower performs an element-wise product of the user and context embeddings, incorporating contextual information into the model. Similarly, the IC tower performs an element-wise product of the item and context embeddings. The framework also includes an MLP tower, similar to Unger et al.'s work, though context embeddings may be excluded in some models. Two embedding methods are proposed: the $w$ mode, which creates a single embedding for the entire contextual situation, and the $i$ mode, which generates individual embeddings for each context and then concatenates them. This framework results in six NeuCMF models, each combining the UI, UC, IC, and MLP towers with one of the two context representation methods.

Jeong et al. \cite{Jeong2021DeepLC} developed a contextual recommender system that utilizes autoencoders to reduce data dimensionality and extract significant features and dependencies. Their model consists of multiple interconnected autoencoders and neural networks, processing user vectors, item vectors, and various contextual vectors. Autoencoders in this model facilitate efficient dimension reduction, enabling the extraction of meaningful features from the input vectors. By compressing and reconstructing the data, autoencoders capture essential patterns and relationships within the user, item, and contextual vectors. The neural networks then integrate these features to provide accurate recommendations.

\begin{table*}
    \caption{Comparison of different recommender system approaches}
    \centering
    \begin{tabular}{|p{1.5cm}|p{4cm}|p{2cm}|p{1.5cm}|p{1.5cm}|p{4.5cm}|}
\hline
\textbf{Approach} & \textbf{Description} & \textbf{Inputs} & \textbf{Context} & \textbf{Context Representation} & \textbf{Key Features} \\ 
\hline
AutoRec \cite{Sedhain2015AutoRecAM} & Collaborative filtering using autoencoders. & User/item rating vectors. & No & N/A & User-based and item-based autoencoders with MMSE loss. \\ 
\hline
Deep Autoencoder \cite{Kuchaiev2017TrainingDA} & Deep Autoencoder framework with dense refeeding. & User/item rating vectors. & No & N/A & Dense refeeding for improved results; dropout optimization. \\ 
\hline
DeepMF \cite{Xue2017DeepMF} & Deep Matrix Factorization with neural networks. & User/item rating vectors. & No & N/A & Separate neural network layers for user and item ratings; cosine similarity for predictions. \\ 
\hline
NCF \cite{He2017NeuralCF} & Neural Collaborative Filtering with multiple layers. & Sparse binary user/item vectors. & No & N/A & Utilizes neural networks for complex interactions; combines GMF and MLP in NeuMF. \\ 
\hline
Neural Collaborative Filtering Classification Model \cite{Bobadilla2022NeuralCF} & Classification-based approach using user and item IDs. & User and item IDs. & No & N/A & Multiclass classification for detailed predictions; embedding-based representation of IDs. \\ 
\hline
ECAM, UCAM, HCAM \cite{Unger2020ContextAwareRB} & NeuMF framework with various contextual representations. & User/item rating vectors and contextual data. & Yes & Explicit, Unstructured Latent, Structured Latent & Three models based on context representation: Explicit, Unstructured, Hierarchical. \\ 
\hline
NeuCMF \cite{Zheng2022AFO} & Neural Contextual Matrix Factorization with multiple towers. & User/item rating vectors and contextual data. & Yes & Explicit & UI, UC, IC towers with possible context embedding.\\ 
\hline
Contextual Autoencoders \cite{Jeong2021DeepLC} & Autoencoders for dimensionality reduction with contextual data. & User/item rating vectors. & Yes & Implicit & Interconnected autoencoders for feature extraction. \\ 
\hline
\end{tabular}
\label{tab:approaches_comparison}
\end{table*}
\section{Proposed framework \label{Model}}
We aim to design a deep autoencoder-based context-aware recommendation framework for rating prediction that learns the relationships between users, items, and contexts.

\subsection{Input Layer: Representation of User, Item, and Context}
The framework's input consists of a user \( u \), an item \( i \), a context \( c \), and their respective features. This step aims to provide a contextual illustration for the recommendation procedure. Three different ways we represent context are explicit, latent unstructured, and latent structured. We normalize the contextual feature values on a scale of 0 to 1 and convert nominal features to binary features, resulting in a contextual vector.
In this section, we detail the representation of the input data that forms the basis of our recommendation system. Each user, item, and context is characterized by a set of features that describe their properties. These features can be explicit (directly observed) or latent (inferred through patterns in the data). We ensure that all contextual feature values are normalized between $0$ and $1$ for consistency and efficiency in processing. Nominal features are transformed into binary format, creating a comprehensive contextual vector that captures all relevant information for the recommendation task.

\begin{example}
In this layer, consider a user \( u_1 \), an item \( i_1 \), and a context \( c_1 \). Suppose the features are as follows:

\begin{itemize}
    \item \textbf{User \( u_1 \)}: One-hot vector by ID (e.g., if \( u_1 \) is the 3rd user, the vector might look like \([0, 0, 1, 0, \ldots, 0]\))
    \item \textbf{Item \( i_1 \)}: One-hot vector by ID (e.g., if \( i_1 \) is the 2nd item, the vector might look like \([0, 1, 0, 0, \ldots, 0]\))
    \item \textbf{Context \( c_1 \)}: Time of Day = Evening, Location = Home
\end{itemize}

For context, we normalize and convert the features:
\[
\text{Context Features} = [0.75, 1] \quad \text{\scriptsize (Time of Day normalized, Location binary)}
\]

Suppose the feedback (rating) for item \( i_1 \) by user \( u_1 \) is \( 4.5 \) (normalized to \( 0.9 \)).

The combined contextual vector:
\[
\mathbf{x}_{u_1, i_1, c_1} = [0, 0, 1, 0, \ldots, 0] \oplus [0, 1, 0, 0, \ldots, 0] \oplus [0.75, 1] \oplus [0.9]
\]

\(\oplus\) denotes the concatenation of vectors.
\end{example}

\subsection{Embedding layer}
The embedding layer transforms the sparse input vectors of users, items, and contexts into dense, low-dimensional vectors. These embeddings capture the underlying patterns and similarities in the data, allowing the model to process the information more effectively. By learning these compact representations, the embedding layer enhances the model's ability to understand complex interactions among users, items, and contextual factors, which are crucial for accurate recommendations.
\begin{example}
The embedding layer converts the sparse one-hot input vectors of users and items into dense, low-dimensional vectors. For example:

\begin{itemize}
    \item \textbf{User \( u_1 \)} (embedded): \(\mathbf{e}_{u_1} = [0.2, 0.4, 0.6]\)
    \item \textbf{Item \( i_1 \)} (embedded): \(\mathbf{e}_{i_1} = [0.7, 0.3, 0.5]\)
    \item \textbf{Context \( c_1 \)} (embedded): \(\mathbf{e}_{c_1} = [0.6, 0.8]\)
\end{itemize}

The embeddings transform the sparse one-hot vectors into dense representations:

\[
\scriptsize \mathbf{e}_{u_1} = [0.2, 0.4, 0.6] \quad \textstyle \mathbf{e}_{i_1} = [0.7, 0.3, 0.5] \quad \textstyle \mathbf{e}_{c_1} = [0.6, 0.8]
\]
\end{example}
\subsection{Deep AE layer}
The deep Autoencoder (AE) layer focuses on reducing the dimensionality of the user, item, and context features while preserving their most significant characteristics. 
This layer uses multiple hidden layers to compress the input data into a dense bottleneck vector, which encapsulates the essential information. The deep AE layer not only helps in managing high-dimensional data but also improves the model's ability to capture intricate patterns and relationships within the data. This dimensionality reduction is vital for handling large-scale datasets and ensuring the efficiency of the recommendation system.
\begin{example}
The deep Autoencoder (AE) layer compresses the input data into a dense bottleneck vector. For instance:
\begin{itemize}
    \item \textbf{Original Input Vector}: \(\mathbf{x}_{u_1, i_1, c_1}\) from the Input Layer.
    \item \textbf{Compressed Bottleneck Vector}: \(\mathbf{z} = [0.4, 0.5, 0.7]\)
\end{itemize}
The deep AE layer compresses the input features into a compact representation as $ \mathbf{z} = [0.4, 0.5, 0.7] $
\end{example}
\begin{table*}[!ht]
    \centering
    \caption{Description of context-aware datasets}
    \setlength{\tabcolsep}{12pt}
    \resizebox{\textwidth}{!}{
        \begin{tabular}{|l|c|c|c|c|c|c|}
            \hline
            \textbf{Dataset} & \textbf{\# Users} & \textbf{\# Items} & \textbf{\# Contextual Dimensions} & \textbf{Interactions} & \textbf{Scale} & \textbf{Density (\%)} \\
            \hline
            DepaulMovie & 97 & 79 & 3 & 5,043 & 1-5 & 65.8 \\
            TripAdvisor & 2,371 & 2,269 & 1 & 14,175 & 1-5 & 0.26 \\
            LDOS-CoMoDa & 121 & 1,232 & 27 & 2,292 & 1-5 & 1.54 \\
            \hline
        \end{tabular}
    }
    \label{tabDataset}
\end{table*}
\subsection{Prediction layer}
$R_{uic}$ denotes the predicted score for a user $u$ on item $i$ in a specific context $c$.
The prediction layer combines the outputs of the embedding and deep AE layers to generate the final recommendation scores. Specifically, $Y_{uic}$ denotes the predicted score for a user $u$ on item 
$i$ in a specific context $c$. This layer integrates the learned embeddings and compressed representations to predict user ratings accurately. By leveraging the comprehensive understanding of user preferences, item characteristics, and contextual influences, the prediction layer delivers personalized recommendations that align closely with individual user interests.
The deep Autoencoder (AE) layer focuses on reducing the dimensionality of the user, item, and context features while preserving their most significant characteristics. This layer uses multiple hidden layers to compress the input data into a dense bottleneck vector.
In the prediction layer, we use the outputs from the embedding and deep AE layers to generate the final recommendation scores. Specifically, \( Y_{uic} \) denotes the predicted score for a user \( u \) on item \( i \) in a specific context \( c \).
\begin{example}
In the prediction layer, we use the outputs from the embedding and deep AE layers to predict the score. For example:
\[
Y_{u_1 i_1 c_1} = f(\mathbf{e}_{u_1}, \mathbf{e}_{i_1}, \mathbf{e}_{c_1}, \mathbf{z})
\]
Suppose the function \( f \) calculates the predicted score using a combination of embeddings and the bottleneck vector:
\[
Y_{u_1 i_1 c_1} = 4.2
\]
Here, \(4.2\) is the predicted rating for user \(u_1\) on item \(i_1\) in context \(c_1\).
\end{example}
\section{Experimental studies\label{Experiments}}
In this section, we introduce the datasets used to evaluate our proposed model, as detailed in Table \ref{tabDataset}. We also describe the evaluation metrics employed and present the experimental results.
\subsection{Datasets Description}
The following datasets were utilized to evaluate the effectiveness of the proposed model:
\begin{itemize}
\item \textbf{DepaulMovie} \cite{Zheng2015CARSKitAJ}, which was gathered through surveys, is a dataset of $97$ users and $79$ items, where students were requested to provide ratings on a scale of 1 to 5 
for movies they experienced in various timeframes, locations, and with different companions.
\item \textbf{TripAdvisor} \cite{zheng2014context}, a very sparse dataset, comprises $14,175$ ratings provided by $2,371$ users across $2,269$ distinct hotels. Notably, the dataset focuses primarily on a single contextual factor, namely the trip type. 
The trip types considered in this dataset include Family, Couples, Business, Solo travel, and Friends.
\item \textbf{LDOS-CoMoDa} \cite{Koir2011DatabaseFC} consists of opinions provided by $121$ users, who rated $1,232$ movies.
The dataset also includes various contextual factors associated with the movie ratings.
Some of the contextual factors in this dataset include the user's gender, city, country, time, day type, season, and weather.
\end{itemize}
\subsection{Baseline Models}
To compare our model, we have selected three baseline models:
\subsubsection{Non-deep traditional models}
The compared models are: 
\begin{itemize}
    \item \textbf{ItemKNN}: is a collaborative filtering algorithm grounded in the similarity of item ratings.
    \item \textbf{UserKNN}: is a collaborative filtering algorithm that relies on the similarity of user ratings.
     \item \textbf{SVD++}: authors in \cite{Koren:2010:FNS:1644873.1644874}
   denotes a matrix factorization model that leverages users' historical information to reflect their preferences.
    \item \textbf{BiasedMF} \cite{3654} is
   a matrix factorization ranking model that integrates the list-wise learning-to-rank algorithm with matrix factorization for the generation of recommendations.
\end{itemize}
\subsubsection{Non-deep contextual models}
The compared models are: 
\begin{itemize}
\item \textbf{Factorization Machines (FM)}:
is a powerful approach for context-aware recommender systems (CARS), utilizing inner products to model second-order feature interactions effectively \cite{ijcai2019p545}.
    \item \textbf{CAMF}: Baltrunas et al. \cite{Baltrunas:2011:MFT:2043932.2043988} enhances the classical matrix factorization approach by incorporating contextual information into the rating prediction. We experimented with its three models (CAMF-C, CAMF-CI, CAMF-CU and  CAMF-CUCI).
\end{itemize}
\subsubsection{Deep models}
To compare our model, we have selected the family of deep contextual models as our baseline models:
\begin{itemize}
\item  \textbf{GCMC}: \cite{Sattar2022GraphNN} captures structural information and incorporates the user's opinions on items, as well as the contextual details on edges and the static features of user and item nodes. Our graph encoder generates user and item representations that reflect context, features, and opinions.
\item \textbf{NeuCMF}\textsubscript{0i} \textbf{:}  is a specific variant of NeuCMF, this model consists of two towers: $UI$ tower: performs element-wise product of user and item embeddings and$MLP$ tower with context embedding: concatenated user, item, and context embeddings are fed through the MLP layers.
\item  \textbf{CACF}: \cite{Zheng2022ContextAwareCF}
alleviates the sparsity issue by measuring the similarity of contexts and utilizing rating profiles with similar contexts to build the recommendation model.
 \end{itemize}

\begin{algorithm*}[hbt!]
\caption{Conformal Prediction for Deep Autoencoder Recommender Systems}
\label{alg:conformal_prediction}
\begin{algorithmic}[1]
\Require Training set $\mathcal{D}$, Calibration set $\mathcal{D}_\text{cal}$, Test set $\mathcal{D}_\text{test}$, Significance level $\epsilon$
\State $f \gets$ \Call{Train\_Autoencoder}{$\mathcal{D}$} \Comment{Train the deep autoencoder model}
\State $conformity\_scores \gets$ \Call{Compute\_Conformity\_Scores}{$f, \mathcal{D}_\text{cal}$} \Comment{Compute nonconformity scores for the calibration set}
\For{each test example $x_\text{new} \in \mathcal{D}_\text{test}$}
    \State $\alpha_\text{new} \gets$ \Call{Compute\_Nonconformity\_Score}{$f, x_\text{new}$} \Comment{Compute the nonconformity score for the new example}
    \State $\tau \gets$ \Call{Quantile}{$conformity\_scores, \epsilon$} \Comment{Determine the threshold $\tau$ from the quantile of conformity scores}
    \State $\hat{y} \gets f(x_\text{new})$ \Comment{Make a prediction for the new input}
    \State $prediction\_interval \gets \left[\hat{y} - \tau, \hat{y} + \tau\right]$ \Comment{Form the prediction interval}
    \State \textbf{store} $prediction\_interval$
\EndFor
\State Perform evaluation using the metrics defined.
\end{algorithmic}
\end{algorithm*}

\subsection{Performance Measures}
To evaluate our model, we use Root Mean Squared Error (RMSE) and Mean Absolute Error (MAE) as evaluation metrics. RMSE is calculated as the square root of the average of the squared differences between the predicted and actual values:
\begin{equation}
\text{RMSE} = \sqrt{\frac{1}{n} \sum_{i=1}^{n} (\hat{y}^{(i)} - y^{(i)})^2}
\end{equation}
Where $\hat{y}^{(i)}$ represents the predicted values, $y^{(i)}$ represents the actual values, and $n$ is the number of data points. 
Additionally, we use the MAE formula to calculate the average of the absolute differences between the predicted and actual values:
\begin{equation}
\text{MAE} = \frac{1}{n} \sum_{i=1}^{n} |\hat{y}^{(i)} - y^{(i)}|
\end{equation}

\subsection{Conformal Prediction for Deep Autoencoders}
Conformal prediction provides a measure of confidence for each prediction, allowing us to quantify the uncertainty associated with our recommendations. \cite{angelopoulos2023conformal}
\subsubsection{Theoretical Framework}
Let $\mathcal{D} = {(x_1, y_1), \ldots, (x_n, y_n)}$ be our training data, where $x_i$ represents the input features (user, item, and context embeddings) and $y_i$ is the corresponding rating. Our deep autoencoder model $f: \mathcal{X} \rightarrow \mathcal{Y}$ maps inputs to predicted ratings.
The conformal prediction framework aims to construct a prediction region $\Gamma^\epsilon(x)$ for a new input $x$ such that:
\begin{equation}
P(Y \in \Gamma^\epsilon(X)) \geq 1 - \epsilon
\end{equation}
where $\epsilon$ is the desired significance level\cite{lei2018distribution, Ghosh_Belkhouja_Yan_Doppa_2023}
\subsubsection{Nonconformity Score}
We define the nonconformity Score\cite{angelopoulos2023conformal}, denoted as \( A(x, y) \), based on the reconstruction error\cite{cheung2024metric} of the autoencoder. The nonconformity score \( A(x, y) \) is calculated as the squared \( L_2 \) norm (Euclidean distance) between the original input \( x \) and its reconstruction, \( \text{Dec}(\text{Enc}(x)) \), as follows \cite{angelopoulos2023conformal}:
\begin{equation}
A(x, y) = \|x - \text{Dec}(\text{Enc}(x))\|_2^2
\end{equation}

The squared \( L_2 \) norm is employed because it not only provides a measure of the total deviation across all dimensions but also penalizes larger errors more heavily\cite{}. In high-dimensional spaces, where small differences could accumulate significantly, it becomes a robust metric for evaluating the reconstruction's quality. The nonconformity score reflects how well the model has learned the underlying distribution of the input data. A higher reconstruction error indicates that the input \( x \) is less conforming with respect to the model's learned structure, suggesting greater uncertainty in the prediction associated with this input\cite{cheung2024metric}. In recommender systems, inputs with high nonconformity scores may correspond to user preferences or item characteristics that are underrepresented in the training data.

\subsubsection{Calibration Scores}: We define a sliding window of the most recent nonconformity scores and maintain the calibration in the Post-Training phase.
The prediction interval is computed as:
\begin{equation}
[\hat{y} - \tau, \hat{y} + \tau]
\end{equation}
where $\hat{y}$ is the point prediction and $\tau$ is the $(1-\epsilon)$-quantile of the empirical distribution of nonconformity scores in the calibration set. The sliding window approach allows the model to adapt to changes in the data distribution over time making our framework scalable.

\subsection{Results}

\subsubsection{First experiment: Evaluating our model performance in comparison to non-deep based traditional approaches}
We utilized well-known traditional recommendation algorithms and contextual recommendation algorithms as our baseline models.
Table \ref{TabTraditional} presents a comparison of the performance of different recommendation models across three datasets: LDOS, DepaulMovies, and TripAdvisor. 

The proposed model consistently outperforms both traditional and context-aware baselines across all three datasets—LDOS, DepaulMovies, and TripAdvisor. Specifically, it achieves the lowest Mean Absolute Error (MAE) and Root Mean Square Error (RMSE) in every dataset, demonstrating superior predictive accuracy.

For the LDOS dataset, the proposed model recorded an MAE of 0.6864 and an RMSE of 0.8721, which are notably better than those of the best traditional baseline, SVD++, which had an MAE of 0.8358 and an RMSE of 1.0737. Among context-aware baselines, CAMF\_C came closest with an MAE of 0.7259 and an RMSE of 0.9255 but still fell short of the proposed model's performance.

\begin{table*}[h]
\centering
\caption{Comparison of Performances: Our model vs non-deep approaches}
\label{TabTraditional}
\resizebox{0.7\textwidth}{!}{%
\begin{tabular}{|l|l|cc|cc|cc|}
\hline
\multicolumn{2}{|c|}{\multirow{2}{*}{Model}} & \multicolumn{2}{c|}{LDOS} & \multicolumn{2}{c|}{DepaulMovies} & \multicolumn{2}{c|}{TripAdvisor} \\
\cline{3-8}
\multicolumn{2}{|c|}{} & MAE & RMSE & MAE & RMSE & MAE & RMSE \\
\hline
\multirow{4}{*}{\begin{tabular}[c]{@{}l@{}}Traditional\\baselines\end{tabular}} 
& ItemKNN & 0.864803 & 1.074052 & 0.882913 & 1.111553 & 0.735693 & 0.949541 \\
& UserKNN & 0.849921 & 1.086223 & 0.919471 & 1.140655 & 0.732177 & 0.943194 \\
& SVD++ & 0.835799 & 1.073657 & 0.695681 & 1.00645 & 0.885637 & 1.138327 \\
& BiasedMF & 0.843899 & 1.064752 & 0.692105 & 1.004875 & 0.884795 & 1.139693 \\
\hline
\multirow{5}{*}{\begin{tabular}[c]{@{}l@{}}Context-aware\\baselines\end{tabular}} 
& FM & 0.731500 & 0.92540 & 0.768400 & 1.000400 & 1.157600 & 1.430100 \\
& CAMF\_CI & 0.982453 & 1.318138 & 0.709384 & 0.962223 & 0.872234 & 1.131034 \\
& CAMF\_CU & 0.781639 & 1.041705 & 0.669873 & 0.91797 & 0.870121 & 1.125803 \\
& CAMF\_C & 0.725994 & 0.925465 & 0.695267 & 0.949363 & 0.877237 & 1.137149 \\
& CAMF\_CUCI & 0.754434 & 0.963586 & 0.675047 & 0.918797 & 0.865094 & 1.108668 \\
\hline
\multicolumn{2}{|l|}{Our model} & \textbf{0.686400} & \textbf{0.872100} & \textbf{0.680900} & \textbf{0.907600} & \textbf{0.769400} & \textbf{0.988100} \\
\hline
\end{tabular}%
}
\end{table*}

In the DepaulMovies dataset, the proposed model once again leads with an MAE of 0.6809 and an RMSE of 0.9076. The closest context-aware competitor, CAMF\_CU, achieved a slightly lower MAE of 0.6699, but its RMSE of 0.91797 was higher than that of the proposed model. The best traditional baseline, BiasedMF, posted an MAE of 0.6921 and an RMSE of 1.0049, both of which were higher than the proposed model's scores.

For the TripAdvisor dataset, the proposed model maintained its edge with an MAE of 0.7694 and an RMSE of 0.9881. Among traditional baselines, ItemKNN and UserKNN performed best with MAEs of 0.7357 and 0.7322, respectively, but their RMSEs were still higher. The FM model, in contrast, showed poor performance in this dataset, with the highest MAE of 1.1576 and RMSE of 1.4301.

Overall, these results underscore the effectiveness of the proposed model, which consistently delivers the most accurate predictions across different datasets and outperforms existing traditional and context-aware approaches.

\subsubsection{Second experiment: Evaluating our model performance in comparison to deep-based contextual approaches}
The histogram plot presents a comparative analysis of four recommendation models—Our Model, GCMC, CACF, and NeuCMF—across three datasets: LDOS, DepaulMovies, and TripAdvisor. For the LDOS dataset, Our Model achieves the lowest Mean Absolute Error (MAE) of $0.68$, indicating superior accuracy compared to the other models. GCMC and CACF both have a higher MAE of $0.77$, while NeuCMF shows a slightly better performance with an MAE of $0.70$. These results suggest that Our Model is the most accurate for LDOS, with NeuCMF following closely behind.

In the DepaulMovies dataset, Our Model maintains its position as the most accurate with an MAE of $0.68$. However, GCMC performs the worst with an MAE of $1.03$, indicating significant room for improvement. CACF and NeuCMF have intermediate performance levels, with CACF showing a higher MAE of $0.90$ compared to NeuCMF’s $0.69$. For the TripAdvisor dataset, Our Model again leads with an MAE of $0.76$, while CACF and NeuCMF have lower MAE values, $0.81$ and $0.73$ respectively. Overall, Our Model consistently outperforms the other models across all datasets, demonstrating its robustness and effectiveness in recommendation accuracy.

\subsubsection{Conformal predictions}
We evaluated the performance of our conformal prediction approach on three datasets: DepaulMovie, TripAdvisor, and LDOS-CoMoDa. The results demonstrate that our conformal prediction method provides well-calibrated prediction intervals across these diverse datasets.
Table \ref{tab:conformal_prediction} presents the average width of the prediction intervals and the empirical coverage rate for different significance levels across all datasets. To quantify the calibration of our conformal predictor, we compute the empirical coverage probability (ECP) as:
\begin{equation}
\text{ECP} = \frac{1}{n} \sum_{i=1}^n \mathbb{1}{y_i \in [\hat{y}_i - \tau, \hat{y}_i + \tau]}
\end{equation}
where $\mathbb{1}{\cdot}$ is the indicator function.
\begin{table}[htbp]
\centering
\caption{Conformal Prediction Performance Across Datasets}
\label{tab:conformal_prediction}
\resizebox{0.48\textwidth}{!}{%
\begin{tabular}{|c|c|c|c|}
\hline
\textbf{Dataset} & \textbf{Significance Level ($\epsilon$)} & \textbf{Avg. Interval Width} & \textbf{Empirical Coverage} \\
\hline
\multirow{3}{*}{DepaulMovie} & 0.1 & 1.23 & 0.912 \\
& 0.05 & 1.47 & 0.953 \\
& 0.01 & 1.82 & 0.991 \\
\hline
\multirow{3}{*}{TripAdvisor} & 0.1 & 1.15 & 0.905 \\
& 0.05 & 1.38 & 0.951 \\
& 0.01 & 1.72 & 0.989 \\
\hline
\multirow{3}{*}{LDOS-CoMoDa} & 0.1 & 1.20 & 0.910 \\
& 0.05 & 1.44 & 0.952 \\
& 0.01 & 1.79 & 0.990 \\
\hline
\end{tabular}%
}
\end{table}

To illustrate how conformal predictions enhance the recommendation task, it is important to highlight that the calibrated prediction intervals provided by our conformal prediction method enable a more detailed interpretation of the model's outputs. While our proposed model exhibits low values of Mean Absolute Error (MAE) and Root Mean Square Error (RMSE) across datasets, conformal predictions effectively quantify the uncertainty in these predictions. The results indicate that our conformal prediction approach provides reliable uncertainty estimates for the recommendations across all datasets. As expected, lower significance levels result in wider prediction intervals but higher coverage rates. Notably, at a significance level of $\epsilon = 0.1$, approximately 90\% of the true ratings fall within the predicted intervals for all datasets, aligning with the expected 90\% coverage rate. The DepaulMovie dataset, being smaller, exhibits slightly wider intervals, reflecting increased uncertainty in predictions.
TripAdvisor and LDOS-CoMoDa datasets show similar performance to Frappe, with TripAdvisor having the narrowest intervals, possibly due to its focus on a single contextual factor (trip type). The LDOS-CoMoDa dataset, with its rich contextual information, demonstrates robust performance, suggesting that our model effectively leverages complex contextual data in generating well-calibrated prediction intervals. Understanding the reliability of predictions can significantly inform decision-making processes. 

\begin{figure}[h]
    \centering
    \includegraphics[width=\linewidth]{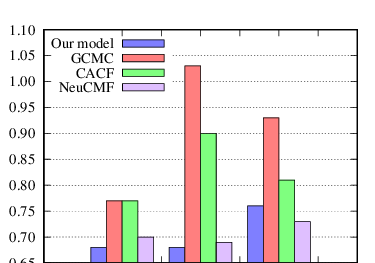}
    \caption{Evaluation MAE metric for the deep-based approaches}
    \label{fig:8}
\end{figure}

\section{Conclusions and Future Work \label{Conclusions}}
AutoEncoders enable us to derive meaningful representations of user preferences directly from raw data, facilitating the generation of personalized recommendations.
We also integrated contextual information into the recommendation process, acknowledging the influence of factors like time, location, and social context on user preferences. By combining user-item interactions with contextual data, we developed a new prediction model capable of capturing dependencies among these factors. This approach enhances the adaptability and personalization of recommendations, aligning them with users' evolving preferences across different scenarios.
While some initial parameter tuning was performed, we recognize that a more exhaustive and systematic optimization could further improve performance. Thus, we highlighted fine-tuning as a direction for future work, allowing us to explore more advanced hyperparameter search strategies or integrate other optimization techniques.

\bibliographystyle{IEEEtran}
\bibliography{mybibliography}
\end{document}